\documentclass[journal,twoside,web]{ieeecolor}
\usepackage{generic}
\usepackage{amsmath,amssymb,amsfonts}
\usepackage{algorithmic}
\usepackage{graphicx}
\usepackage{algorithm}
\usepackage{hyperref}
\hypersetup{hidelinks=true}
\usepackage{textcomp}
\usepackage{cite}
\usepackage{caption}
\usepackage{booktabs}

\usepackage{xcolor}         
\usepackage{colortbl}  
\definecolor{gray94}{gray}{.94}
\definecolor{gray90}{gray}{.90}
\newcommand{\grow}[1]{\rowcolor{gray94}{#1}}

\newtheorem{proposition}{Proposition}

\def\BibTeX{{\rm B\kern-.05em{\sc i\kern-.025em b}\kern-.08em
    T\kern-.1667em\lower.7ex\hbox{E}\kern-.125emX}}
\begin{document}

\title{\textit{D-Flow}: Multimodal Flow Matching for D-peptide Design}
\author{Fang Wu, Shuting Jin, Xiangru Tang, Junlin Xu, Mark Gerstein, Li Erran Li, \emph{Fellow, IEEE}, and James Zou 
\thanks{Fang Wu and James Zou are with the Department of Computer Science, Stanford University, Palo Alto, CA 94305, USA (e-mails: fangwu97@stanford.edu; jamesz@stanford.edu).}%
\thanks{Shuting Jin and Junlin Xu are with the School of Computer Science and Technology, Wuhan University of Science and Technology, Wuhan 430065, China (e-mails: shutingjin@wust.edu.cn; xjl@wust.edu.cn).}%
\thanks{Xiangru Tang and Mark Gerstein are with the Department of Computer Science, Yale University, New Haven, CT 06520, USA (e-mails: xiangru.tang@yale.edu; mark.gerstein@yale.edu).}%
\thanks{Li Erran Li is with AWS AI, Amazon, Santa Clara, CA 94305, USA (e-mail: lilimam@amazon.com).}%
\thanks{This work was supported in part by the National Natural Science Foundation of China (Grant No.62402351); the Hubei Provincial Natural Science Foundation of China (Grant No.2024AFB275); the Scientific Research Project of Education Department of Hubei Province (Grant No.Q20231109) (Corresponding author: Shuting Jin).}%
}
\maketitle
\begin{abstract}
Proteins are crucial to biological processes, and therapeutic peptides are emerging as promising pharmaceutical agents. 
Among these, D-peptides are resistant to proteolysis, exhibit greater in vivo stability, and are easier to synthesize. Despite advances in deep learning for peptide discovery, the scarcity of natural D-protein data limits the transfer of existing generative models to the D-peptide chemical space. 
We propose D-Flow, a full-atom flow-based framework for \emph{de novo} D-peptide design. Conditioned on receptor binding, D-Flow uses structural representations incorporating backbone frames, side-chain angles, and discrete amino acid types. A mirror-image algorithm is implemented to address the lack of training data for D-proteins by converting the chirality of L-receptors. Furthermore, we enhance D-Flow's capacity by integrating protein language models (PLMs) with structural awareness through a lightweight structural adapter that injects structural representations into PLM embeddings. This enables D-Flow to learn conformational priors in the D-peptide chemical space and to accommodate the chiral selectivity of binding sites, thereby mitigating the scarcity of D-peptide data.
A two-stage training pipeline and a control toolkit enable D-Flow to transition from general protein design to targeted binder design while preserving pre-training knowledge. Results on the PepMerge benchmark show D-Flow's effectiveness. D-peptides generated by D-Flow align more closely with native sequences and structures, with sequence identity improving by 10.2\% over the best baseline and the top affinity score reaching 24.31\%.
{Overall, D-Flow shows potential for D-peptide design, facilitating the development of bioorthogonal and stable molecular tools and diagnostics. Code is available at \url{https://github.com/smiles724/PeptideDesign}.}
\end{abstract}
\begin{IEEEkeywords}
D-peptide design, Peptide drugs, Multimodal, Flow matching
\end{IEEEkeywords}

\maketitle
\section{Introduction}
\IEEEPARstart{P}{roteins} are the building blocks of life and play essential roles in nearly all biological processes. Remarkably, therapeutic peptides, comprising a limited number of well-ordered residues, are single-chain proteins and an irregular class of pharmaceutical agents~\cite{wang2022therapeutic}. Peptide drugs occupy a unique chemical and pharmacological space between small and large molecules~\cite{lin2024ppflow}. Notably, unbound peptide chains exhibit high free energy and entropy, leading to unstable conformations. In contrast, they can elicit pharmacological effects upon binding to specific receptors, forming stable complexes. Given this specialty, an increasing number of studies adopt deep learning (DL) algorithms to facilitate peptide discovery~\cite{li2024full}. 

Flow models~\cite{lipman2022flow} are revolutionizing several domains, including drug discovery~\cite{wu2023molformer,wu2024hierarchical}. They have been applied to various protein representations, including carbon-alpha only structures~\cite{trippe2022diffusion}, torsion angles~\cite{wu2024protein}, and the SE(3) backbone frame representation~\cite{yim2023se}, and diverse scenarios such as molecular design~\cite{guan20233d}, antibody engineering~\cite{martinkus2024abdiffuser}, \emph{de novo} protein design~\cite{yim2023fast}, and peptide discovery~\cite{ramasubramanian2024hybrid} as well.
\begin{figure*}[t]
    \centering
    \includegraphics[width=1\textwidth]{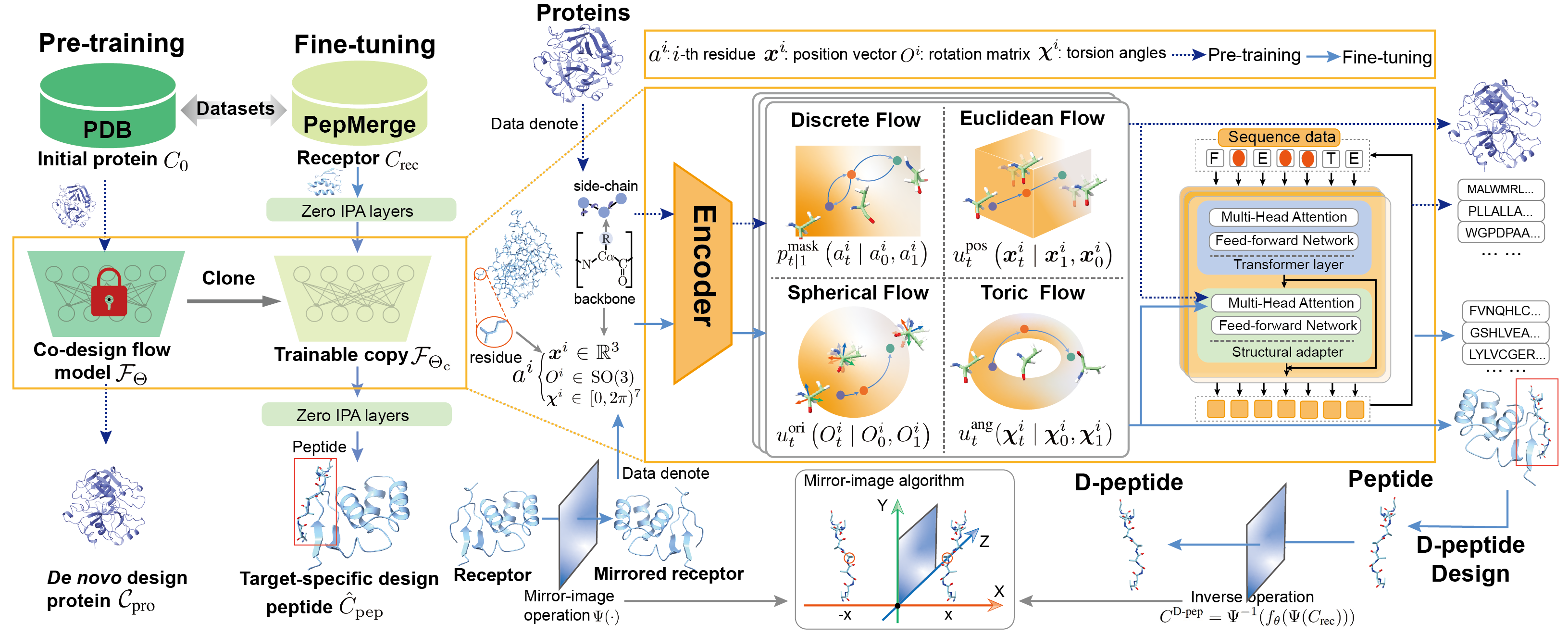}
    \caption{Illustration of D-Flow. The co-design flow is pre-trained on PDB, and a trainable copy is leveraged for subsequent fine-tuning on the downstream peptide design. To produce full-atom structures, the core flow model comprises four modalities: discrete residue types, the Euclidean space of alpha-carbon coordinates, the spherical space of backbone frames, and the toric space of side-chain angles. Finally, designed peptide sequences are fed into the frozen protein language model for refinement via a structural adapter. During inference, D-Flow produces D-peptides using a mirror-image algorithm, where a mirrored D-receptor protein is used as the model input, and the generated peptide structure is mirrored back as the final output.}
    \label{fig:model}
\end{figure*}
Despite growing interest in peptide design, the generation of D-proteins remains underexplored. Though every cell in the human body contains proteins, their cornerstone, 20 categories of amino acids, can exist in two stereoisomers: L (levo) and D (dextro), which are non-superimposable mirror images of each other~\cite{lander2023d}, and this chirality characteristic is determined by the orientation of the functional groups around the alpha carbon. D-proteins are protein molecules whose polypeptide chains consist of D-amino acids and the achiral amino acid glycine. They can form specific heterochiral protein-protein interactions (PPIs) with natural L-protein targets and possess remarkable potential as therapeutics and diagnostics~\cite{sun2024accurate}. 
However, proteins made entirely of D-amino acids have not been observed in nature, and many natural peptides contain only a few D residues, such as bacterial non-ribosomal products or certain animal toxins. Although chemical synthesis is feasible, mechanistically well-characterized D-peptides remain uncommon. Prior generative models typically operate in SE(3) residue frames, handle torsions implicitly or via rotamer libraries, and are trained on L-centric datasets with limited transfer to D-peptides. The scarcity of D-peptide training data further constrains the development and application of deep-learning approaches for D-peptide design.

To overcome this obstacle, we introduce a mirror-imaged Multimodal model D-Flow (see Fig.~\ref{fig:model}). D-Flow characterizes the peptide using rigid backbone frames within the SE(3) manifold, side-chain angles on the high-dimensional torus, and discrete residue types on the categorical space (see Supplementary Information S.1 for details). Each modality has an analytical flow, and they jointly capture peptides' full-atom structures. During inference, a mirror-image algorithm transforms the L-receptor and converts its natural chirality. Unlike existing peptide design frameworks that require task-specific training data, we mitigate the need for D-peptide training corpus by leveraging geometric parity symmetry.
Moreover, as datasets containing complete 3D structures of peptide complexes are orders of magnitude smaller than sequence databases, the scarcity of high-quality peptide-receptor pairs inevitably constrains the performance of DL approaches. As a resolution, we empower the flow model with large protein language models (PLMs) and a two-stage training pipeline. Specifically, structural surgery is applied to PLMs using a lightweight structural adapter~\cite{zheng2023structure}, which endows them with structural awareness and facilitates regression to residue categories from intermediate-state conformational information. Moreover, we employ a control toolkit~\cite{zhang2023adding} to achieve awareness of target proteins and to transfer from general protein design to the binder design task, while incurring minimal loss of pretraining knowledge.  
Experiments on PepMerge~\cite{li2024full} verify the effectiveness of D-Flow, particularly for its breakthrough in generating pure D-peptides. It also outpaces existing DL baselines in generating L-peptides by a large margin. Our study opens avenues for systematic exploration of the DL-based mirror-image protein universe, paving the way for a wide range of design applications targeting L-proteins. 

\section{Related Works}
\label{Appendix:sec:RelatedWork}

\subsection{Diffusion and Flow for Protein Design}
{Diffusion models promise to design novel protein structures with custom functions~\cite{wu20213d,wu2023diffmd,wudiffantiseq,wu2026dynamics,wu2025generalized,wu2022pre,wu2022discovering,wu2023instructbio,wu2024surface,jiang2025posex,deng2025predicting}.} Among distinct protein representations, the frame representation has achieved state-of-the-art performance in \emph{de novo} backbone design and motif scaffolds, as exemplified by RFdiffusion~\cite{watson2023novo}, which extends RoseTTAFold ~\cite{krishna2024generalized} with diffusion to generate protein structures tailored to target functions. {Flow methods~\cite{lipman2022flow} offer a deterministic approach by eliminating stochasticity in sampling.} The linear interpolation schedule for flow yields more direct sampling trajectories, thereby reducing the number of integration steps~\cite{lipman2022flow}. 

{Recent developments in DL algorithms for peptides. PepFlow~\cite{li2024full} explicitly models side-chain torsional dynamics using Conditional Flow Matching (CFM). In contrast, PPFlow~\cite{lin2024ppflow} relies on RDE~\cite{luo2023rotamer} to complete full-atom design.} Kong et al.~\cite{kong2024full} present a full-atom diffusion on the latent geometric space learned by VAE. Nevertheless, none considers extending sequence models (\emph{e.g.}, PLMs) and multi-stage training paradigms to address the shortage of available binding structures. By integrating a mirror-image algorithm with multimodal modeling and PLMs, D-Flow transfers receptor-aware design knowledge from L-protein datasets to D-peptides without requiring D-peptide training data.

\subsection{D-peptide Technology}
Proteins are built from chiral molecules, where amino acids exist in two mirror-image forms: L (Levorotatory) and D (Dextrorotatory). In D-peptides, every chiral center at the C$\alpha$ position has the \textbf{D} configuration (i.e., the noncanonical enantiomer) and no L residues remain. Natural ribosomes exclusively use L-amino acids to synthesize proteins, resulting in L-proteins throughout nature. While engineered ribosomes~\cite{goto2008initiating}, post-translational modification systems~\cite{kobayashi2013characterization}, and non-ribosomal peptide synthetases~\cite{ogasawara2018peptide} can incorporate some D-amino acids into L-protein chains, proteins made entirely of D-amino acids have never been found in nature~\cite{ullah2024extended}. They must be produced by total chemical synthesis with iterative screening or selection to enhance stability and target-guided evolution.

Despite the synthetic challenges, D-proteins are valuable research tools. They fold into mirror images of their L-counterparts and offer unique opportunities for studying fundamental protein mechanisms and developing stable molecular binding agents~\cite{dong2024recent}. 
To identify D-proteins capable of binding to a target L-protein, mirror-image peptide phage display methods have been developed~\cite{callahan2024mirror,qi2024mirror}, which involve screening a phage library of L-peptides with a target D-protein. However, it remains challenging to precisely target a specific surface region of the target protein and confirm the presence of valid binders within the initial random library~\cite{sun2024accurate}. More importantly, no prior work has explored DL co-design methods~\cite{luo2022antigen,yim2023se} for D-protein generation, let alone the incorporation and verification of this mirror-image technique.
Physics-based computational approaches such as structure-based docking and Rosetta inverse folding \cite{juraszek2025novo}, together with DL models originally developed for L systems, have the potential to be adapted to D by mirroring structures or redefining torsion-angle parameters.
Yet the paucity of native D-protein data and the limited explicit handling of chirality restrict the transfer to D-peptides. D-Flow addresses these gaps by coupling a mirror-image scheme for paired supervision.

\section{Preliminary and Background}
\textbf{Proteins.} A protein has $n$ residues, each defined by its type, backbone frame, and side-chain torsion angles. The type of the $i$-th residue, denoted by $a^i \in \mathcal{A}=\{1 \ldots 20\}$, is determined by its side-chain R group. The rigid frame is constructed from the coordinates of four backbone heavy atoms $\mathrm{N}$, $\mathrm{C}\alpha$, $\mathrm{C}$, and $\mathrm{O}$, with $\mathrm{C}\alpha$ positioned at the origin. This frame is represented by a position vector $\boldsymbol{x}^i \in \mathbb{R}^3$ and a rotation matrix $O^i \in \mathrm{SO}(3)$. Seven torsion angles $\boldsymbol{\chi}^i \in[0,2 \pi)^7$ are considered and three of them belong to the backbone, where $\boldsymbol{\chi}^i[0]$ is the angle around the $\mathrm{C}-\mathrm{N}$ bond, $\boldsymbol{\chi}^i[1]$ is the angle around the $\mathrm{N}-\mathrm{C}\alpha$ bond, and $\boldsymbol{\chi}^i[2]$ affects the position of the oxygen atom. The side-chain conformation is more flexible than the backbone, involving up to four rotatable torsion angles between side-chain atoms, denoted by $\boldsymbol{\chi}^i[3:] \in[0,2 \pi)^4$. 

\noindent\textbf{Flow on Riemannian Manifolds.} Flow matching (FM) is a simulation-free method for learning a continuous normalizing flow (CNF). On a manifold $\mathcal{M}$ with a Riemannian metric $g$, the CNF $\Phi: \mathcal{M} \rightarrow \mathcal{M}$ is defined by a one-parameter diffeomorphism that integrates along a time-dependent vector field $u_t \in \mathcal{U}$, where $u_t(x)\in\mathcal{T}_x \mathcal{M}$ falls at the tangent space of the manifold at $x\in \mathcal{M}$. 
With an initial condition of $\phi_0(x)=x$, the time-dependent flow $\phi_t: \mathcal{M} \rightarrow \mathcal{M}$ and the final diffeomorphism are attained by setting $\Phi(x)=\phi_1(x)$ and solving the ordinary differential equation (ODE) $\frac{\mathrm{d} \phi_t}{\mathrm{d}t} ({x}) = u_t(\phi_t(x))$ on $\mathcal{M}$ over $t\in [0,1]$. Remarkably, $\phi_t(x)$ transports the point $x$ along the vector field $u_t(x)$ from time $0$ up to time $t$, satisfying another ODE: $\mathrm{d} x= u_t(x) \mathrm{d}t$. 
Denote $\mathbb{P}(\mathcal{M})$ as the space of probability distributions on $\mathcal{M}$. 

The density $p_t$ is characterized by the Fokker-Planck equation: $\frac{\mathrm{d} p_t}{\mathrm{d} t} = -\textrm{div}(u_t p_t)$, also known as the continuity equation. Under these conditions,  $u_t$ is said to be the probability flow for $p_t$, and $p_t$ is said to be the marginal probability path generated by $u_t$. Although $u_t$ is intractable in general, it can be learned efficiently by decomposing the target probability path $p_t$ as a mixture of tractable conditional probability paths $p_t(x|x_1)$, which satisfy $p_0(x|x_1) = p_0(x)$ and $p_1(x|x_1) = \delta_{x_1}(\cdot)$. 

Conditional Riemannian flow matching (CRFM) fits a parameterized vector field $v_\theta\in \mathcal{U}$ to $u_t(x|x_1)$, which produces $p_t(x|x_1)$. The objective is on the tangent space $\mathcal{T}_x \mathcal{M}$:
\begin{equation}
    \mathcal{L}_{\textrm{CRFM}}(\theta) = \mathbb{E}_{t\sim \mathcal{U}[0,1], p_1(x_1), p_t(x|x_1)} \|v_\theta(x, t) - u_t(x|x_1) \|^2_g,
\end{equation}
this loss can be parameterized by defining the conditional flow $x_t = \phi_t(x_0|x_1)$, where $\phi_t$ is the solution to $\frac{\mathrm{d} \phi_t}{\mathrm{d}t} ({x}) = u_t(\phi_t(x_0|x_1)|x_1)$ with an initial condition of $\phi_0(x_0|x_1) = x_0$. Finally, the loss can be written as $\mathcal{L}_{\textrm{CRFM}}(\theta) = \mathbb{E}_{t\sim \mathcal{U}[0,1], x_1\sim p_1(x_1), x_0\sim p_0(x_0)} \|v_\theta(x, t) - \Dot{x}_t \|^2_g$. During inference, samples can be generated by solving the ODE associated with $v_\theta$ to efficiently move $x_0\in \mathcal{M}$ from the source $p_0$ to the data distribution $p_1$.

\section{Methods}
\subsection{Flow for Full-atom Protein Generation}
\textbf{Discrete Flow for 1D Amino Acid Sequence.} Two recent lines adapt diffusion or flow to the discrete setting: one embeds the discrete data in continuous space~\cite{li2024full}, and the other designs the transformation process immediately over categorical state spaces~\cite{campbell2024generative,gat2024discrete}. We investigate both and observe a significant advantage of the discrete flow over the continuous version.  

We define discrete FM via a Continuous-Time Discrete Markov Chain (CTMC). The categorical variable $a^i_t$ jumps between states in the amino acid type space $\mathcal{A}$ depending on a continuous time $t \in[0,1]$. The \emph{probability velocity} $u_t$ is defined as the rate of probability change of the current sample $a_t^i$ in each of its $\|\mathcal{A}\|=20$ categories. Thus, each category of the residue type $a^i_t\sim p_t$ is updated independently with the Euler step as $a_{t+\frac{1}{N}}^i \sim \delta_{a_t^i}(\cdot)+\frac{1}{N} u_t\left(a^i,\cdot\right)$. For a sufficiently large number of timesteps $N\rightarrow\infty$ and any potential state $z\in\mathcal{A}$, the probability velocity $u_t$ is required to satisfy 
\begin{equation}
    \sum_{a^i \in\mathcal{A}} u_t\left(a^i, z\right)=0 \text{ and } u_t\left(a^i, z\right) \geq 0, 
\end{equation}
for $\forall i \in[n]$ and $a^i \neq z$, and $u_t$ can be constructed as the marginalization of the conditional one $u_t\left(a^i, z \mid a^i_0, a^i_1\right)$. Then, we attain the marginal velocity written as $u_t\left(a^i, z\right)=\sum_{a^i_0, a^i_1 \in \mathcal{A}} u_t\left(a^i, z \mid a^i_0, a^i_1\right) p_t\left(a^i_0, a^i_1\mid z\right)$, which generates the probability path $p_t\left(a^i\right)$. 
Generally, the conditional probability paths can be represented as a convex sum of $m$ conditional probabilities $p_t\left(a^i\mid a^i_0, a^i_1\right) = \sum_{j=1}^k \kappa_t^j p^j(a^i\mid a^i_0, a^i_1)$, where the schedulers $\kappa_t^j$ are collectively non-negative and satisfy $\sum_j \kappa_t^j = 1$. 
A simple yet useful instance is $p_t\left(a^i\mid a^i_0, a^i_1\right) = \kappa_t \delta_{a^i_1}\left(a^i\right) + (1- \kappa_t) \delta_{a^i_0}\left(a^i\right)$, where the scheduler $\kappa_t$ satisfies $\kappa_0 = 0$, $\kappa_1=1$ and monotonically increases in $t$ (\emph{i.e.}, $\dot{\kappa}_t\geq 0$).  
This results in $p_0(a^i | a^i_0, a^i_1)= \delta_{a^i_0}(a^i)$ and $p_1(a^i|a^i_0, a^i_1)=\delta_{a^i_1}(a^i)$. Subsequently, we get the marginal probability of velocity:
\begin{equation}
    u_t\left(a^i, z\right)=\frac{\dot{\kappa}_t}{1-\kappa_t}\left[p_{1 \mid t}\left(a^i \mid z\right)-\delta_z\left(a^i\right)\right],
\end{equation}
where $p_{1 \mid t}\left(a^i \mid z\right)=\sum_{a_0^i, a_1^i} \delta_{a_1^i}\left(a^i\right) p_t\left(a_0^i, a_1^i \mid z\right)$ denotes the probability denoiser. The training goal is to learn $p_{1 \mid t}$ by minimizing the cross-entropy (CE) loss:
\begin{equation}
\begin{split}
    \mathcal{L}_{\text{aa}}(\theta)=&\mathbb{E}_{i\in[n], t, p_0\left(a^i\right),p_1\left(a^i\right), p_{t}\left(a^i | a^i_0, a^i_1\right)}
    \log p_{1 \mid t}\left(a^i_1 \mid a^i_t\right),
\end{split}
\end{equation}
where $t$ is uniformly sampled on $[0,1]$, and a network $v^{\text{aa}}\left(a^i_t,t\right)$ with parameters $\theta$ approximates $p_{1\mid t}$. Given a noisy input  $a_t^i \sim p_{t\mid 1}\left(a^i\mid a_1^i, a_0^i\right)$, the model learns to predict the clean data $a_1^i$. 
Here, rather than a linear interpolation towards $a_1^i$ from a prior of a uniform categorical distribution $p_{t\mid 1}^{\text{unif}}\left(a_t^i \mid a_0^i,a_1^i\right) = \operatorname{Cat}\left(t\delta_{a_1^i}(\cdot) + (1-t) \frac{1}{\|\mathcal{A}\|}\right)$, we adopt a mask state $\text{M}$ and the conditional path becomes:
\begin{equation}
\label{equ: p_t_1}
    p_{t \mid 1}^{\text{mask}}\left(a_t^i \mid a_0^i,a_1^i\right) = \text{Cat}\left(t\delta_{a_1^i}(\cdot) + (1-t)\delta_{\text{M}}(\cdot)\right).
\end{equation}
A mask-state prior focuses probability on plausible states, while a uniform prior wastes mass on irrelevant ones, leading to more efficient and accurate inference.

\noindent\textbf{Multimodal FM for 3D Protein Structures.} We construct different probability flows containing Euclidean, spherical, and toric FMs for positions $\boldsymbol{x} \in \mathbb{R}^{n\times 3}$, orientations $O\subseteq \mathrm{SO}(3)$, and torsion angles $\boldsymbol{\chi} \in [0, 2\pi)^{n\times 7}$, respectively. To be specific, a vanilla Gaussian FM on Euclidean manifolds with the prior $\mathcal{N}\left(0, \boldsymbol{I}_3\right)$ is leveraged to generate $\boldsymbol{x}^i$. The 3D rotation group $\mathrm{SO}(3)$ is also a smooth Riemannian manifold with its tangent space $\mathfrak{so}(3)$ forming a Lie algebra consisting of skew-symmetric matrices. We establish flows based on the geodesics in $\mathrm{SO}(3)$ and select the uniform distribution over $\mathrm{SO}(3)$ as the prior $p\left(O_0^i\right)$. For torsion angles $\boldsymbol{\chi}^i\in[0, 2\pi)^7$, each can be represented as a point on the unit circle $\mathbb{S}^1$. Thus, $\boldsymbol{\chi}^i$ lies on the 7-dimensional hypertorus $\mathbb{T}^7 = (\mathbb{S}^1)^7$ as the Cartesian product of all seven unit circles. This flat torus $\mathbb{T}^7$ can be viewed as the quotient space $ (\mathbb{R}^7 \mod 2\pi\mathbb{Z} )^7 $ that inherits the Riemannian metric from Euclidean space, where the uniform distribution on $[0, 2\pi)^7$ is the prior. The conditional flows are
\begin{align}
    \phi_t^{\text{pos}}\left(\boldsymbol{x}_0^i, \boldsymbol{x}_1^i\right)&=t \boldsymbol{x}_1^i+(1-t) \boldsymbol{x}_0^i, \quad \boldsymbol{x}_0^i\sim \mathcal{N}\left(0, \boldsymbol{I}_3\right), \\
    \phi_t^{\text{ori}}\left(O_0^i, O_1^i\right) & =\exp_{O_0^i}\left(t \log _{O_0^i}\left(O_1^i\right)\right), \quad O_0^i \sim \mathcal{U}(\mathrm{SO}(3)), \\ 
    \phi_t^{\text{ang}}(\boldsymbol{\chi}_0^i, \boldsymbol{\chi}_1^i) &= \left[t \boldsymbol{\chi}_1^i+(1-t)\boldsymbol{\chi}_0^i\right]\text{mod} 2\pi,  \boldsymbol{\chi}_0^i \sim \mathcal{U}\left([0, 2\pi)^7\right), 
\end{align}
where $\text{exp}(\cdot)$ and $\text{log}(\cdot)$ are the exponential and logarithm maps on $\mathrm{SO}(3)$ that can be computed efficiently using Rodrigues' formula. Subsequently, $u_t^{\text{pos}}$, $u_t^{\text{ori}}$, and $u_t^{\text{ang}}$ can be obtained by taking the time derivative of linear flows $\phi_t^{\text{pos}}$, $\phi_t^{\text{ori}}$, and $\phi_t^{\text{ang}}$ using Independent Coupling techniques:
\begin{align}
    u_t^{\text{pos}}\left(\boldsymbol{x}_t^i \mid \boldsymbol{x}_1^i, \boldsymbol{x}_0^i\right)&=\boldsymbol{x}_1^i-\boldsymbol{x}_0^i=\frac{\boldsymbol{x}_1^i-\boldsymbol{x}_t^i}{1-t}. \\
    u_t^{\text{ori}}\left(O_t^i \mid O_0^i, O_1^i\right) & =\frac{\log_{O_t^i} \left(O_1^i\right)}{1-t},\\
    u_t^{\text{ang}}(\boldsymbol{\chi}_t^i \mid \boldsymbol{\chi}_0^i, \boldsymbol{\chi}_1^i ) &= \left(\frac{\boldsymbol{\chi}_1^i - \boldsymbol{\chi}_t^i}{1-t} + \pi \right)\mod 2\pi - \pi, 
\end{align}
Finally, time-dependent and equivariant networks $v^{\text{pos}}(\cdot)$, $v^{\text{ori}}(\cdot)$, and $v^{\text{ang}}(\cdot)$, which share the same backbone architecture as $v^{\text{aa}}(\cdot)$ but have different head predictors, are used to approximate $u_t^{\text{pos}}$, $u_t^{\text{ori}}$, and $u_t^{\text{ang}}$. The FM objectives are: 
\label{equ: loss_pos}
\begin{equation}
\begin{split}
    \mathcal{L}_{\text{pos}}(\theta)=&\mathbb{E}_{i\in [n],t\sim \mathcal{U}(0,1), p\left(\boldsymbol{x}_1^i\right), p\left(\boldsymbol{x}_0^i\right), p\left(\boldsymbol{x}_t^i|\boldsymbol{x}_0^i,\boldsymbol{x}_1^i\right)}\\
    & \left\|v^{\text{pos}}\left(\boldsymbol{x}_t^i, t\right)-\left(\boldsymbol{x}_1^i-\boldsymbol{x}_0^i\right)\right\|_2^2, \\    
\end{split}    
\end{equation}
\begin{equation}
\begin{split}
    \mathcal{L}_{\text{ori}}(\theta)=&\mathbb{E}_{i\in [n],t\sim \mathcal{U}(0,1), p\left(O_1^i\right), p\left(O_0^i\right),p\left(O_t^i|O_0^i,O_1^i\right)}\\
    &\left\|v^{\text{ori}}\left(O_t^i, t\right)-\frac{\log _{O_t^i} \left(O_1^i\right)}{1-t}\right\|_{\mathrm{SO}(3)}^2, \\
\end{split}
\end{equation}
\begin{equation}
\begin{split}
    \mathcal{L}_{\text{ang}}(\theta)=& \mathbb{E}_{i\in [n],t\sim \mathcal{U}(0,1), p\left(\boldsymbol{\chi}_1^i\right), p\left(\boldsymbol{\chi}_0^i\right),p\left(\boldsymbol{\chi}_t^i|\boldsymbol{\chi}_0^i,\boldsymbol{\chi}_1^i\right)}\\
    & \left\|v^{\text{ang}}\left(\boldsymbol{\chi}_t^i, t\right)-  (\boldsymbol{\chi}_1^i - \boldsymbol{\chi}_0^i)\right\|_2^2.  
\end{split}
\end{equation}
Together with the sequence flow, these geometric flows enable D-Flow to model amino acid identities, backbone geometries, and side-chain conformations within a unified framework rather than in separate stages. As a result, dependencies between sequence and structure can be maintained during generation, which improves the geometric and chemical consistency of the designed peptide binders. These objectives are then combined into $\mathcal{L}_{\mathrm{CFM}} = \lambda_{\text{pos}} \mathcal{L}_{\text{pos}} + \lambda_{\mathrm{ori}} \mathcal{L}_{\text{ori}} + \lambda_{\text{ang}} \mathcal{L}_{\text{ang}} + \lambda_{\text{aa}}  \mathcal{L}_{\text{aa}}$, where $\lambda_*$ are the hyperparameters to control loss components. Two additional losses are also imposed concerning the backbone atoms and the distance matrix~\cite{yim2023fast}. 

During inference, we first sample from several distinct priors, \emph{i.e.}, $\boldsymbol{x}_0^i \sim$ $\mathcal{N}\left(0, \boldsymbol{I}_3\right)$, $O_0^i \sim \mathcal{U}(\mathrm{SO}(3))$, and $\boldsymbol{\chi}_0^i \sim [0, 2\pi)^7$. After that, we solve the probability flow with $v^{\text{pos}}(\cdot)$, $v^{\text{ori}}(\cdot)$, and $v^{\text{ang}}(\cdot)$ using the $N$-step forward Euler method to get the position, orientations, and torsion angles of the $i$-th residue with $t=\left\{0, \ldots, \frac{N-1}{N}\right\}$:
\begin{align}
    \boldsymbol{x}_{t+\frac{1}{N}}^i&=\boldsymbol{x}_t^i+\frac{1}{N} v^{\text{pos}}\left(\boldsymbol{x}_t^i, t\right), \\
    O_{t+\frac{1}{N}}^i&=\exp_{O_t^i}\left(\frac{1}{N} v^{\text{ori}}\left(O_t^i, t\right)\right), \\
    \boldsymbol{\chi}_{t+\frac{1}{N}}^i&=\left[\boldsymbol{\chi}_t+\frac{1}{N} v^{\text{ang}}\left(\boldsymbol{\chi}_t^i, t\right)\right] \mod 2\pi.
\end{align}

\subsection{Parameterization with Adapter-guided Protein Language Models}
To efficiently learn $(a^i, \boldsymbol{x}^i, O^i, \boldsymbol{\chi}^i)$ for every residue, we build upon FramePred~\cite{yim2023se,yim2023fast}, which incorporates Invariant Point Attention (IPA)~\cite{jumper2021highly} to encode spatial features and ensure equivariance. In addition, considering the periodicity, $\boldsymbol{\chi}\in[0, 2\pi)^7$ are flexibly encoded by applying multi-frequency sine and cosine transformations, fused with the timestep embedding and residue sequence embedding into IPAs.

PLMs capture the evolutionary patterns from large-scale sequence data, and this knowledge is supportive of protein folding~\cite{lin2022language}, inverse design~\cite{zheng2023structure}, and our co-design task. Prior studies~\cite{wu2023integration} immediately append the geometric networks to PLMs. Drawing inspiration from LM-Design~\cite{zheng2023structure}, we employ a lightweight structural adapter to endow PLMs with structural awareness. 
Denoting IPA output as $\boldsymbol{h}_{\text{IPA}}\in \mathbb{R}^{\psi_{\text{IPA}}}$, the $l$-th layer's attention is computed as:
$\boldsymbol{o} = \mathrm{softmax}\left(\frac{\boldsymbol{h}_{\text{seq}}\boldsymbol{W}_Q \cdot \boldsymbol{h}_{\text{IPA}}^\top\boldsymbol{W}_K^\top}{\sqrt{\psi_{\text{seq}}}}\right) \boldsymbol{h}_{\text{IPA}}\boldsymbol{W}_V$,
where $\boldsymbol{h}_{\text{seq}}\in \mathbb{R}^{\psi_{\text{seq}}}$ is the sequential embedding of the last $(l-1)$-th layer. $\boldsymbol{W}_K$, $\boldsymbol{W}_Q$, and $\boldsymbol{W}_V$ are trainable weights for key, query, and value, separately. This allows the usage of sequence- and structure-based information for protein design, enabling PLM to incorporate geometric signals from intermediate structural states. Consequently, the adapter guides residue prediction using both sequence context and structural information, which improves the geometric consistency between predicted amino acid identities and the evolving backbone conformations during training. The training mechanism of the structural adapter is described in Supplementary Information Section S.2.

\subsection{Controlling Flow Matching with Target Proteins}
Recent studies~\cite{luo2022antigen,li2024full,lin2024ppflow} on binder design are strictly restricted by the number of available complex structures (\emph{e.g.}, SabDAB~\cite{dunbar2014sabdab} and PepMerge~\cite{li2024full}). To bridge the gap and exploit all crystal structures, we propose to pretrain our flow model on the vast amount of general proteins (\emph{i.e.}, Protein Data Bank) before target-specific fine-tuning.

Let $\mathcal{F}_{\Theta}:\mathcal{P}_0 \rightarrow \mathcal{C}_{\text{pro}}$ be a co-design flow model without target awareness, which can \emph{de novo} design a protein ${C}_{\text{pro}}$ from any initial protein ${C}_{0}$ that is drawn from a prior $p_0$. It is trained on general proteins to approximate $p(a, \boldsymbol{x}, O, \boldsymbol{\chi})$. To enable awareness of $\mathcal{F}_\Theta$ to the receptor $C_{\text{rec}}$, we leverage the ControlNet technique~\cite{zhang2023adding}, locking (freezing) the parameters $\Theta$ of the original block and simultaneously cloning the block to a trainable copy with parameters $\Theta_c$. The trainable copy takes a receptor protein $C_{\text{rec}}$ as input and is connected to the locked model with zero IPAs denoted as $\mathcal{Z}(\cdot)$, whose weights and bias are initialized to zeros. In practice, two instances of zero IPAs are used with parameters $\Theta_{\mathrm{z}1}$ and $\Theta_{\mathrm{z}2}$, respectively:
\begin{equation}
    \hat{C}_{\mathrm{pep}}=\mathcal{F}_{\Theta}(C_{0})+\mathcal{Z}_{\Theta_{\mathrm{z} 2}}\left(\mathcal{F}_{\Theta_{\mathrm{c}}}\left(C_{0} +\mathcal{Z}_{\Theta_{\mathrm{z} 1}}\left(C_{\text{rec}}\right) \right)\right),
\end{equation}
where $\hat{C}_{\mathrm{pep}}$ is the designed peptide that is expected to bind with $C_{\text{rec}}$. In the first training step, since all parameters of IPAs are initialized to zero, both $\mathcal{Z}(\cdot)$ terms evaluate to zero, and $\hat{C}_{\mathrm{pep}}=\hat{C}_{\mathrm{pro}}$.
This ControlNet-style conditioning introduces receptor information through a trainable branch while keeping the pretrained model frozen. The zero-initialized IPAs ensure that receptor signals are injected gradually during training, allowing the model to adapt to receptor-specific design tasks without disrupting the pretrained structural priors. 
Zero-initialized IPA layers protect the backbone by preventing random noise from propagating as gradients during the initial training stage (see Supplementary Information Section S.3 for details).
\begin{table*}[t]
    \caption{Evaluation of methods in the traditional L-peptide sequence-structure co-design task, where metrics are divided into three main categories. The \textbf{best} and \underline{suboptimal} results are labeled boldly and underlined, respectively. }  
    \centering
    \resizebox{1.6\columnwidth}{!}{%
    \begin{tabular}{c|cccc|cc|cc} \toprule
    & \multicolumn{4}{c|}{ \textbf{Geometry} } & \multicolumn{2}{c|}{ \textbf{Energy} } & \multicolumn{2}{c}{ \textbf{Design} } \\
    & AAR $\% \uparrow$ & RMSD $\text{\AA} \downarrow$ & $\operatorname{SSR} \% \uparrow$ & BSR $\% \uparrow$ & Stb. $\% \uparrow$ & Aff. $\% \uparrow$ & Des. $\% \uparrow$ & Div. $\uparrow$ \\ \midrule
    RFdiffusion~\cite{watson2023novo} & 40.14 & 4.17 & 63.86 & {26.71} & \underline{26.82} & 16.53 & \textbf{78.52} & 0.38 \\
    ProteinGen~\cite{lisanza2023joint} & 45.82 & 4.35 & 29.15 & 24.62 & 23.48 & 13.47 & 71.82 & 0.54 \\
    Diffusion~\cite{luo2022antigen} & 47.04 & 3.28 & 74.89 & 49.83 & 15.34 & 17.13 & 48.54 & {0.57} \\
    PPFlow~\cite{lin2024ppflow} & 48.35  &  3.59 & 68.13 & 25.94 & 15.77 & 12.08 & 46.53 & 0.51 \\
    PepFlow-Bb~\cite{li2024full} &  50.46 & 2.30 & 82.17 & 82.17 & 14.04 & 18.10 & 50.03 & \textbf{0.64} \\ 
    PepFlow-Seq~\cite{li2024full} & \underline{53.25} & {2.21} & \underline{85.22} & {85.19} & 19.20 & 19.39 & 56.04 & 0.50 \\ 
    PepFlow-Ang~\cite{li2024full} & {51.25} & \underline{2.07} & {83.46} & \underline{86.89} & 18.15 & \underline{21.37} & 65.22 & 0.42 \\  \midrule
    \grow{D-Flow} & \textbf{58.69} & \textbf{1.63} & \textbf{89.02} & \textbf{88.47} & \textbf{26.85} & \textbf{24.31} & \underline{75.14} & \underline{0.60} \\ \bottomrule
    \end{tabular}}
    \label{tab:co-design}
\end{table*}

\subsection{D-peptide Design}
D-peptides have garnered attention as potential therapeutic and enzymatic tools due to their resistance to enzymatic digestion by natural-chirality enzymes and their exceptional biostability~\cite{zhang2024mirror}. However, the development of mirror-image biological systems and related applications faces challenges, primarily due to the lack of effective methods for designing mirror-image (D-) proteins~\cite{lander2023d}. 

In this work, we generate D-peptides using the mirror-image algorithm~\cite{sun2024accurate}. 
Since the model is trained exclusively on L-protein datasets, a potential bias may arise toward the conformational statistics and interaction patterns characteristic of L-peptides. The mirror-image algorithm mitigates this issue by transforming the receptor geometry into D-space during inference, allowing the model to generate peptides under mirrored stereochemical constraints while preserving the underlying geometric relationships.
The mirror-image operation $\Psi(\cdot)$ transforms a protein structure into its corresponding mirror image. $\Psi(\cdot)$ inverts proteins' spatial configuration, effectively reflecting them across an imaginary plane while preserving the relative distances and angles between atoms.

We first apply $\Psi(\cdot)$ to the receptor $C_{\text{rec}}$, converting it into its mirror-image counterpart. Then, we pass this mirrored receptor through a flow model $f_\theta(\cdot):\mathcal{C}_{\text{rec}}\rightarrow \mathcal{C}_{\text{pep}}$, which is trained on natural-chirality (L-) proteins, to yield the corresponding peptide. Finally, we apply the inverse operation $\Psi^{-1}(\cdot)$ to convert the designed peptide back to the D-configuration. Formally, the generation process can be expressed as:
\begin{equation}
\label{equ:mirror}
    C^{\text{D-pep}}=\Psi^{-1}(f_\theta (\Psi(C_{\text{rec}}))). 
\end{equation}
Notably, $\Psi(\cdot)$ can be represented as a
parity transformation of the form $\Psi(C)=MC$, where $M$ is a reflection
matrix satisfying $M^\top M=I$ and $\det(M)=-1$. Unlike rotation matrices that preserve orientation with $\det=+1$, a reflection matrix reverses chirality (e.g., converting an L-protein into its D-enantiomer) while preserving all pairwise distances and angles. Because $M$ is an improper rotation, it does not, in general, commute with arbitrary elements of $SO(3)$.

\section{Results}
\label{ExperimentalResults}
We execute two types of experiments to validate the effectiveness of our D-Flow. The first, detailed in Sec.~\ref{sec:lprotein}, involves the conventional co-design challenge for L-proteins, where models co-design peptides conditioned on a given receptor binding site. The second, described in Sec.~\ref{sec:dprotein}, co-designs D-proteins. For benchmarking, we use the \textbf{PepMerge} dataset~\cite{li2024full}, derived from PepBDB~\cite{wen2019pepbdb} and Q-BioLip~\cite{wei2024q}. To ensure a fair comparison with prior work~\cite{li2024full}, we cluster peptide-protein complexes based on 40\% sequence identity using MMseqs2~\cite{steinegger2017mmseqs2}, after removing duplicates and applying empirical filters (\emph{e.g.}, resolution $<$ 4\AA, peptide length between 3 and 25). This process resulted in 8,365 non-redundant complexes across 292 clusters. For consistency in comparison, we use the same test set as Li et al.~\cite{li2024full}, consisting of 10 clusters and 158 complexes. Pretraining data contain monomers with lengths between 60 and 384 with resolution $<5\textup{\AA}$ downloaded from PDB~\cite{berman2000protein} on August 8, 2021, ensuring no data leakage for peptides. The data is then filtered by including proteins with high secondary structure compositions only. Monomers with more than 50\% loops are also removed using DSSP~\cite{kabsch1983dictionary}, resulting in 18,684 proteins. 

\subsection{Unconditioned Sequence-structure Co-design}
\label{sec:lprotein}
\textbf{Baselines.}  {We select two categories of baselines.} The first disregards side-chain conformations and includes approaches such as RFDiffusion and ProteinGen~\cite{lisanza2023joint}. RFDiffusion generates backbones, with sequences predicted afterward using ProteinMPNN~\cite{dauparas2022robust}, while ProteinGen improves on RFDiffusion by jointly sampling both backbones and sequences. {The second considers full-atom protein generation, comprising Diffusion~\cite{luo2022antigen}, PPFlow, and PepFlow, which are recent diffusion- and flow-based models most closely related to D-Flow. See Supplementary Information S.4 for details.} PepFlow has three variants based on whether backbones, sequences, and side-chain angles are sampled. For each test case, we sample 64 peptides simultaneously. 

\noindent\textbf{Metrics.} Peptides are evaluated across three key aspects. (1) \textit{Geometry}: Peptides should closely mirror native sequences and structures. We quantify sequence similarity through the amino acid recovery rate (\textbf{AAR}), which measures sequence identity between generated and ground truth peptides. Structural similarity evaluation uses the root-mean-square deviation (\textbf{RMSD}) of $C_\alpha$ atoms after complex alignment. The secondary-structure similarity ratio (\textbf{SSR}) captures the proportion of matching secondary structures, while the binding site ratio (\textbf{BSR}) evaluates the overlap between generated and native peptide binding sites on the target protein. (2) \textit{Energy}: We aim to design high-affinity peptide binders that enhance protein-peptide complex stability. \textbf{Affinity} represents the percentage of generated peptides achieving higher binding affinities (lower binding energies) than the native one. \textbf{Stability} indicates the fraction of complexes exhibiting lower total energy than the native state. All energy calculations use Rosetta~\cite{alford2017rosetta}.
(3) \textit{Design}: \textbf{Designability} measures the consistency between designed sequences and structures, calculated as the proportion of sequences that fold into structures similar to their generated conformations ($C_\alpha$ RMSD $<$ 2~\textup{\AA}). Sequence refolding is performed using ESMFold~\cite{lin2022language}. \textbf{Diversity}, computed as the average of one minus the pairwise TM-Score~\cite{zhang2004scoring}, quantifies structural variation among the designed peptides. All metrics are under the same protocols across datasets, experiments, and baseline comparisons to ensure consistency and fairness.
\begin{table}[t]
\centering
\caption{Ablation studies on the effects of each module. To be specific, DFM is the abbreviation of \emph{discrete flow matching}, PLM is the usage of \emph{protein language models}, Pretrain is the transfer learning from PDB to PepMerge, and CN is the employment of the \emph{ControlNet} technique. }
\resizebox{1.0\columnwidth}{!}{
\begin{tabular}{@{}c|cccc| cc | cc@{}}
\toprule
 & DFM & PLM & Pretrain & CN & AAR$\% \uparrow$ & $\Delta$ & RMSD$\downarrow$ & $\Delta$ \\ \midrule
1 & - & -  & - &  - & 50.43 &  -- &  2.34  & --  \\
2 & \checkmark & -  & - & - & 51.86  & +1.43  & 2.09 & - 0.25 \\
3 & \checkmark & \checkmark  & - & - & 52.44 & +0.58  &  2.01 & -0.08\\
4 & \checkmark & \checkmark & \checkmark & - & 58.37 & +5.93  &  1.78 & -0.23 \\  \midrule
\grow{5} & \checkmark & \checkmark & \checkmark & \checkmark & 58.69 & +0.32 & 1.63 & -0.15 \\ \bottomrule
\end{tabular}}
\label{tab: ablation}
\end{table}

\noindent\textbf{Main Results.} As documented in Tab.~\ref{tab:co-design}, D-Flow generates peptide sequences with the closest resemblance to native ones, achieving an AAR of 58.69\%, a 14.51\% improvement over PepFlow-Ang. It excels in all geometry-related metrics, highlighting its close alignment with the binding sites of native peptides. Besides, D-Flow scores 75.14\% in designability and 0.60 in diversity, maintaining a good balance between structural fidelity and variety. Moreover, D-Flow also demonstrates superior energy-based properties, achieving the best stability score at 26.85\% and the best affinity score at 24.31\%, critical for the formation of strong and stable complexes. In summary, D-Flow outperforms all baselines across most key metrics, indicating its strength in consistently producing the most accurate peptide sequences and structures, along with the optimal stability and affinity for their targets. 
\begin{figure}
    \centering
    \includegraphics[width=\linewidth]{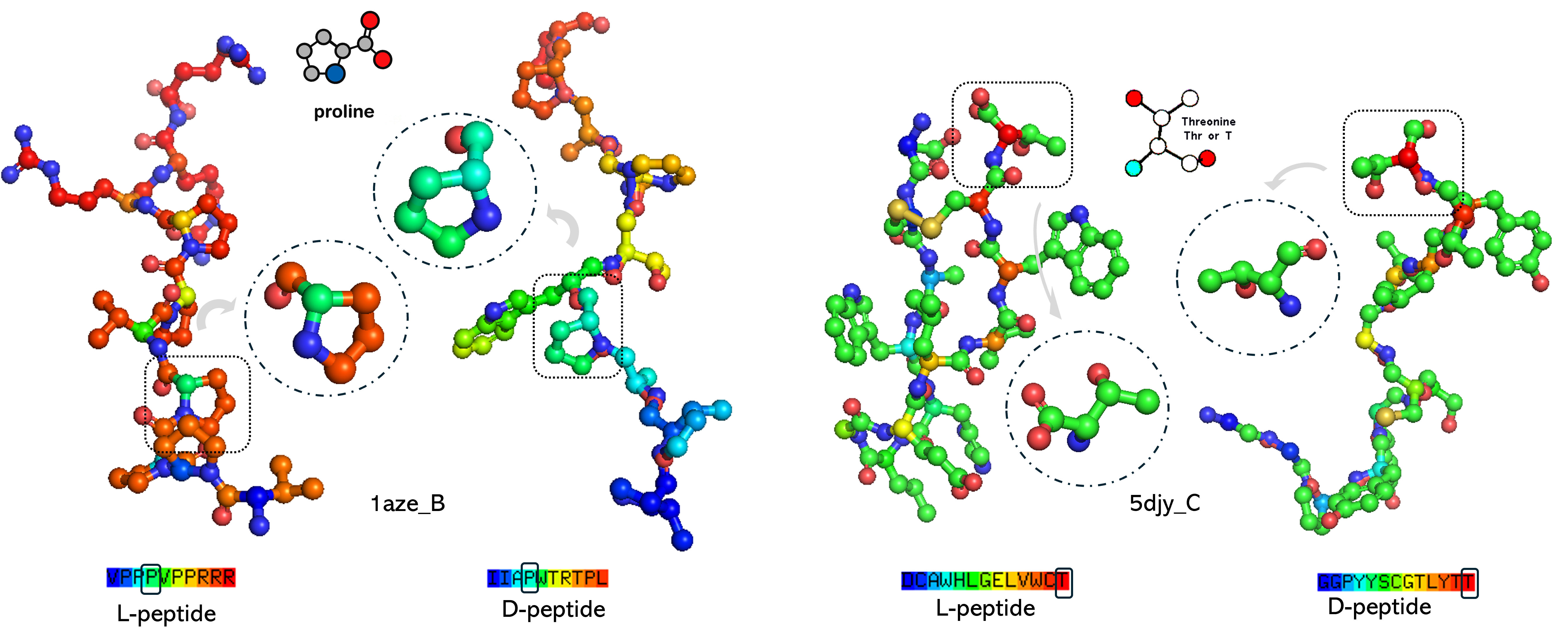}
    \caption{Perspective view of D-residues within two randomly selected D-peptides generated by D-Flow, where all stereogenic alpha carbons to the amino group in D-peptides have D-configuration. }
    \label{fig:d-residue}
\end{figure}
\begin{figure}
    \centering
    \includegraphics[width=0.85\linewidth]{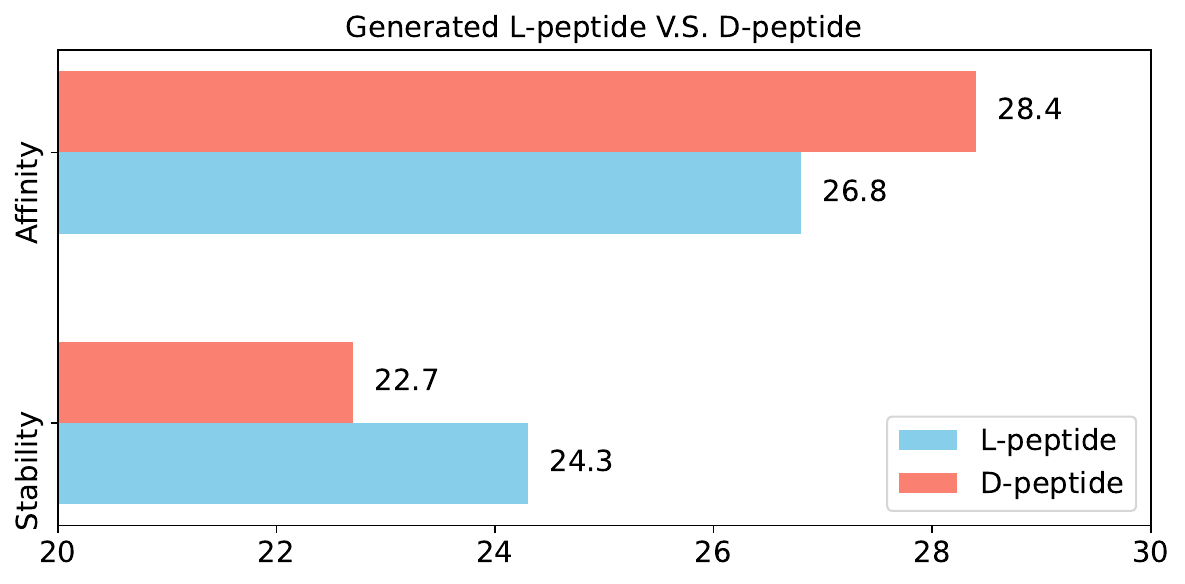}
    \caption{Energy-based comparison between generated L- and D-peptides, simulated by Rosetta. \emph{Stability} represents the fraction of generated complexes exhibiting lower total energy than the native states, while \emph{Affinity} indicates the percentage of generated peptides having higher affinity than the native ones. }
    \label{fig:compare}
\end{figure}

We analyze the contribution of D-Flow components, containing the discrete flow matching (DFM), structurally adapted PLM, pretraining on PDB, and the ControlNet-style transfer learning technique. Tab.~\ref{tab: ablation} reports that the integration of additional unlabeled structural data yielded the most substantial improvement, increasing AAR by 5.93\%. This supports our hypothesis that limited binding complex data significantly constrains generative models' \emph{de novo} design capabilities. 
While incorporating 1D evolutionary information improved D-Flow's performance by 0.58\%, this impact was less pronounced than in previous protein-related tasks like ligand docking~\cite{corso2022diffdock} and ligand efficacy prediction~\cite{townshend2020atom3d}.
We attribute this to PLMs' inherent limitations with short sequences like peptides, compared to their effectiveness with longer proteins. This gap generally arises because PLMs learn contextual relationships by observing long-range dependencies and conserved motifs, key elements that characterize protein families and evolutionary relationships. 
In peptides, these relationships are limited due to a lack of length and structure, making PLMs harder to capture evolutionary patterns effectively. 
For other components, DFM demonstrated superior performance to the probability simplex mechanism for amino acid type representation, improving AAR by 1.43\%. Moreover, the ControlNet-style module improves AAR from 58.37\% to 58.69\% (a 0.32\% increase) and reduces RMSD from 1.78 to 1.63, demonstrating its contribution to receptor-aware peptide design.
\begin{figure*}[t!]
    \centering
    \includegraphics[width=1.0\textwidth]{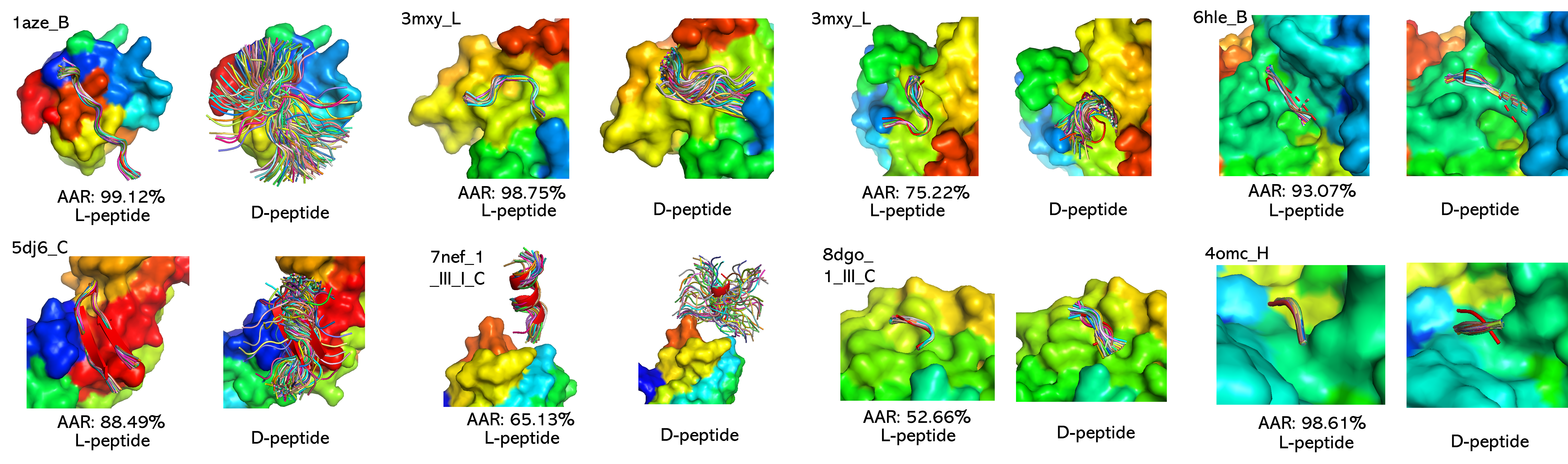}
    \caption{Visualization of L- and D-peptide examples generated by D-Flow with the same target receptor, where the red one is the ground truth L-peptide. Despite that, D-Flow achieves particularly high recovery of both amino acid sequences and structures for L-peptides; its sampling space of D-peptides is relatively diverse and unconstrained. This phenomenon of diversity can be explained by the distribution mismatch between the training and inference data, loss of evolutionary constraints, asymmetric energy landscape exploration, and stereochemical environment adaptation. }
    \label{fig:d-peptide}
\end{figure*}

\textbf{Evaluation Beyond PepMerge.} We expand our evaluation to other large-scale peptide-protein benchmarks to cover substantially more diverse receptor families, peptide lengths, and structural motifs than PepMerge. Specifically, we incorporate the PepBench and ProtFrag datasets for training, comprising: (i) 4,157 experimentally determined protein-peptide complexes; (ii) 70,498 high-quality synthetic complexes generated via AlphaFold-Multimer refinement; and (iii) 114 held-out complexes for validation. For testing, we use the LNR dataset, which contains 93 protein-peptide complexes~\cite{tsaban2022harnessing} spanning peptide lengths of 4-25 residues and diverse binding geometries not present in PepMerge. Together, these datasets provide a substantially broader and more challenging evaluation landscape, mitigating the risk of dataset-specific overfitting.

In Tab.~\ref{tab:lnr_results}, D-Flow achieves strong and consistent performance across all metrics, confirming that it generalizes beyond PepMerge and does not rely on its distributional biases.
\begin{table}[t]
\centering
\caption{{Performance on the LNR peptide-protein benchmark.  C-RMSD: C$\alpha$ RMSD (Å); L-RMSD: ligand RMSD (Å).}}
\label{tab:lnr_results}
\begin{tabular}{lccc}
\toprule
\textbf{Model} & \textbf{AAR} & \textbf{C-RMSD} & \textbf{L-RMSD} \\
\midrule
RFDiffusion   & 34.68\% & 4.69  & 1.88  \\
PepFlow       & 35.47\% & 2.87  & 1.79  \\
PepGLAD       & 38.62\% & 2.74  & 1.60  \\
D-Flow        & \textbf{39.01\%} & \textbf{2.70} & \textbf{1.54} \\
\bottomrule
\end{tabular}
\end{table}

\textbf{Inference Efficiency.} We benchmarked D-Flow, PepFlow, and RFdiffusion on an NVIDIA A100-80GB GPU over the 20 randomly selected protein-peptide complexes, generating 32 peptides per method. Flow models (D-Flow and PepFlow) require only a small number of deterministic transport steps, whereas RFdiffusion performs iterative denoising over rigid-frame parameterizations. As reported in Tab.~\ref{tab:inference_efficiency}, D-Flow achieves $4.1$\,s per peptide, comparable to PepFlow ($4.7$\,s), while RFdiffusion requires $\sim120$\,s per peptide, consistent with the commonly reported $1$-$3$ min runtime even for short peptides. Overall, D-Flow is $25$--$30\times$ faster than RFdiffusion under matched hardware and task settings.
\begin{table}[t]
\centering
\caption{Inference efficiency comparison on an NVIDIA A100-80GB GPU. 
Runtimes are measured for each designed peptide across 20 complexes, yielding 32 designs per method.}
\label{tab:inference_efficiency}
\begin{tabular}{lcc}
\toprule
\textbf{Model} & \textbf{Steps} & \textbf{Runtime / peptide} \\
\midrule
D-Flow & 20 FM steps & 4.1\,s \\
PepFlow & 20 FM steps & 4.7\,s \\
RFdiffusion & 200 DDPM steps & $\sim$120\,s \\
\bottomrule
\end{tabular}
\end{table}

\subsection{D-peptide Design}
\label{sec:dprotein}
\textbf{D-peptides Generation.} Our analysis confirms that D-Flow successfully produces \emph{pure D-peptides} (Fig.~\ref{fig:d-residue}). Remarkably, D-flow requires no D-protein training corpus  but only a specialized post-processing technique $\Psi(\cdot)$.
To understand it, as D-Flow gradually moves the peptide's sequence and structure $(a, \boldsymbol{x}, O, \boldsymbol{\chi})$ from a prior $p_0$ to the target $p_1$ at a speed of $v_t$ during $t\in[0,1]$, its moving trajectory is heavily dependent on the receptor $\mathcal{C}_{\text{rec}}$. Once a D-chirality receptor $\Psi(\mathcal{C}_{\text{rec}})$ is provided as the condition, our D-Flow $f_\theta(\cdot)$ captures its chirality nuance. As a response, it adjusts $v^{\text{pos}}(\cdot), v^{\text{ori}}(\cdot)$, and $v^{\text{ang}}(\cdot)$ to accord with this D-target protein, which leads to a product with considerable D-residues. 

\noindent\textbf{Quantitative Analysis.} Though real-world D-peptides that bind to receptors are unavailable, we simulate via Rosetta to quantitatively analyze the stability and binding affinity for generated D-peptides. As the conformation distribution is divergent, we generate 1K samples for each receptor in the test set and select the most stable one for evaluation. These filtered D-peptides exhibit competitive energy metrics compared to L-peptide counterparts (Fig.~\ref{fig:compare}). They achieve an average improvement in binding affinity of \textbf{28.4\%} over L-peptides, while maintaining a high stability improvement of 22.7\%, significantly outperforming the state-of-the-art L-peptides (24.31\%). Results demonstrate that D-Flow effectively leverages receptor-specific chirality information to optimize peptide-protein interactions.

\noindent\textbf{Model Robustness Assessment.} 
We examined the robustness of D-Flow by testing multiple mirror-image approaches for the input receptor. Despite there being infinite possible mirror transformations, we focused on the three fundamental axes (x, y, and z). Results demonstrate that the chirality of generated peptides remains consistently D-configured regardless of the mirroring approach used. See Supplementary Information S.5 for more robustness assessment.

\noindent\textbf{Conformational Distribution Analysis.}
Fig.~\ref{fig:d-peptide} visualizes designed L- and D-peptides. While D-Flow successfully generates both configurations, we reveal a striking contrast: L-peptides show remarkable consistency in their structural and sequence distributions, whereas D-peptides exhibit significantly higher variability, particularly in longer sequences. This phenomenon can be attributed to several fundamental factors.

\textit{Training Distribution Mismatch.} D-Flow, trained exclusively on L-peptides, has learned to navigate L-peptide conformations' natural distribution and energy landscape. When presented with a D-receptor, the model must operate in a conformational space absent from its training distribution. This shift forces the model to explore the D-space with less confidence, increasing structural diversity.
\textit{Loss of Evolutionary Constraints.} L-peptides reflect millions of years of evolutionary optimization, containing implicit biases about energetically favorable conformations~\cite{levy2020evolutionary}. While the mirror-image transformation preserves theoretical symmetry, these evolutionary constraints don't translate perfectly to the D-space, potentially contributing to the observed variability.
\textit{Asymmetric Energy Landscape Exploration.} FM relies on smooth transitions across the conformational space~\cite{lipman2022flow}. However, the mirror-image transformation introduces subtle numerical asymmetries in the computational representation of D-peptides. These asymmetries can affect how the model explores the energy landscape, leading to more diverse intermediate states and final structures~\cite{wolynes1996symmetry}.
\textit{Stereochemical Environment Adaptation.} The significant change in the stereochemical environment during mirroring poses a unique challenge. D-Flow must adapt its learned representations of molecular interactions to account for the inverted chirality~\cite{anand2022protein}. This adaptation process isn't perfect, resulting in broader conformational sampling and increased structural diversity in D-peptides.

These challenges align with findings that highlight the challenges of transferring learned molecular representations across different stereochemical spaces~\cite{strokach2022deep}. The energy landscape learned for L-peptides doesn't directly translate to the D-space due to fundamental differences in molecular interactions and packing arrangements. This leads to broader structural exploration and increased diversity in the generated structures.
We suggest potential avenues for improvement, such as incorporating symmetry-aware constraints or developing specialized adaptation mechanisms for D-peptide design, to stabilize the output distribution while maintaining the desired chirality.
\begin{figure}[t]
    \centering
    \includegraphics[width=0.6\linewidth]{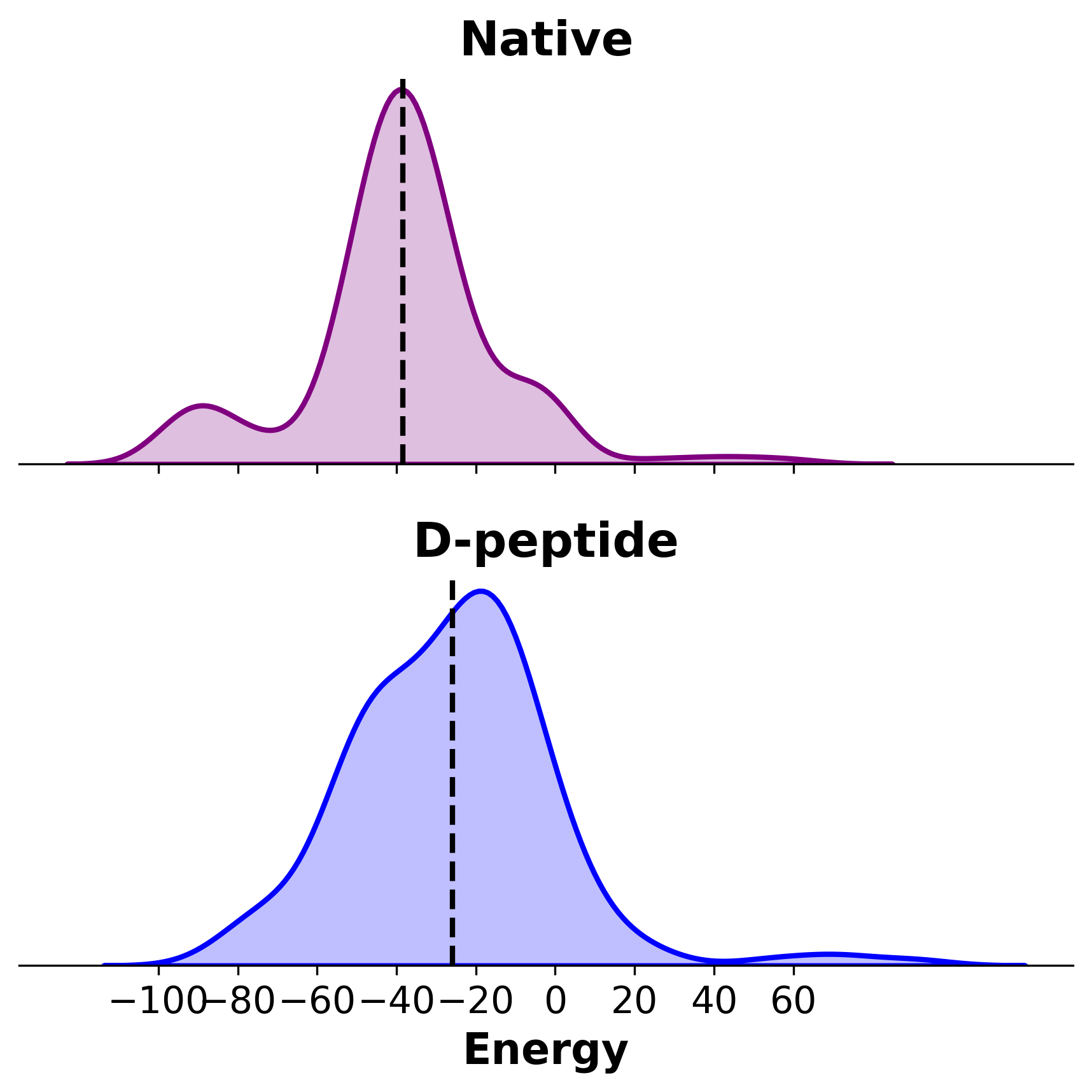}
    \caption{
 \textbf{Comparison of Rosetta energy distributions for generated L- and D-peptides.} Kernel density estimates (KDEs) of Rosetta full-atom energies for (left) generated L-peptides and (right) D-peptides. }
    \label{fig:energy_kde}
    
\end{figure}

\section{Discussion}
\label{Discussion}

\textbf{{Theoretical Property of the Mirror-Image Algorithm.}}
{The reflection $\Psi$ is an isometry of $\mathbb{R}^3$ that preserves all geometric invariants used in standard molecular energy functions. Accordingly, common physics-based potentials (e.g., Rosetta all-atom energy) are parity invariant: for any configuration $C$, $E(\Psi(C)) = E(C)$ when $E$ depends only on bond lengths, angles, torsions, and pairwise distances. Thus, reflecting both receptor and peptide preserves the receptor–ligand energy landscape up to a rigid motion.}

{This symmetry yields two practical consequences for D-peptide design. First, $\Psi(C_{\mathrm{rec}})$ remains a valid conditioning input since its geometry and energetics are unchanged. Second, applying $\Psi^{-1}$ to the generated peptide produces a physically consistent D-configuration, preserving all geometric relations except chirality. As shown in Fig.~3, Rosetta stability and affinity scores remain unchanged under reflection, confirming that the mirror-image algorithm introduces no energetic artifacts. We formalize this symmetry by assuming the model operates on geometric features invariant to both proper and improper rigid motions.}
{\begin{proposition}[Mirror-image Coherence]
\label{prop:mirror_equivariance}
Suppose the generative model $f_\theta$ is equivariant with respect to all 
orthogonal transformations, i.e., $f_\theta(Q C + t) = Q\, f_\theta(C) + t$ for $\forall\, Q \in \mathrm{O}(3),~ t \in \mathbb{R}^3$. Define the mirror-adapted map as $ g_\theta(C) = \Psi^{-1}\!\left( f_\theta\!\left( \Psi(C) \right) \right)$. Then $g_\theta$ is roto-translationally equivariant. In particular, for $\forall\, R \in \mathrm{SO}(3),~ t \in \mathbb{R}^3$, we have
\begin{equation}
    g_\theta(R C_{\mathrm{rec}} + t) = R\, g_\theta(C_{\mathrm{rec}}) + t .
\end{equation}
\end{proposition}}

{\textit{Proof.} Since $\Psi(C) = M C$ with $M \in \mathrm{O}(3)$ and $M^2 = I$, applying a rigid transformation $(R, t)$ before mirroring yields $\Psi(R C_{\mathrm{rec}} + t) = M R C_{\mathrm{rec}} + M t$. Under $\mathrm{O}(3)$-equivariance, $f_\theta(M R C_{\mathrm{rec}} + M t) = M R\, f_\theta(C_{\mathrm{rec}}) + M t$. 
Applying $\Psi^{-1} = M$ and using $M^2 = I$ gives
$\Psi^{-1}\!\left(f_\theta\!\left(\Psi(R C_{\mathrm{rec}} + t)\right)\right)   = R\, f_\theta(C_{\mathrm{rec}}) + t = R\, \Psi^{-1}\!\left(f_\theta(\Psi(C_{\mathrm{rec}}))\right) + t \Psi^{-1}\!\left(f_\theta\!\left(\Psi(R C_{\mathrm{rec}} + t)\right)\right)   = R\, f_\theta(C_{\mathrm{rec}}) + t  = R\, \Psi^{-1}\!\left(f_\theta(\Psi(C_{\mathrm{rec}}))\right) + t$.
Thus, $g_\theta$ inherits SE(3) equivariance in this idealized setting.}

{\textbf{Remark.} Although standard SE(3)-equivariant models are only equivariant to proper rigid motions, $\Psi$ is an isometry that preserves all geometric features used by the model (e.g., distances, angles, and frame-based features). Thus, the mirror-adapted formulation remains geometrically consistent without requiring explicit reflection equivariance.}

{Traditional peptide therapeutics, typically composed of L-amino acids, are rapidly degraded by proteases, limiting their half-life and clinical utility. In contrast, D-peptides resist enzymatic hydrolysis, offering greater stability and oral bioavailability, and enabling the targeting of previously “undruggable” proteins~\cite{lander2023d}.} To address the scarcity of D-peptide–receptor structures, D-Flow incorporates a PLM with a lightweight structural adapter and ControlNet-style conditioning, preserving pretrained knowledge during fine-tuning. The mirror-image algorithm is used only to map receptor geometry into D space for designing shape-complementary binders, not to model full D-protein biophysics. This enables full-atom D-peptide design with stability and affinity comparable to or exceeding L-peptides. On PepMerge, D-Flow outperforms state-of-the-art methods across geometry, energy, and design metrics, achieving higher fidelity, stability, and binding affinity.

However, key limitations remain. The scarcity of natural D-proteins limits generalization, especially under distribution shifts to unseen receptors. Energy analysis shows L-peptides form a narrow, low-energy basin, while D-peptides exhibit broader, higher-variance distributions, indicating less stable conformations under D-Flow. Computational design alone is insufficient for drug-like optimization; integration with physics-based methods (e.g., docking, molecular dynamics, refinement) is needed. Moreover, results are computational only—experimental validation (synthesis, stability, binding, pharmacology) is required. In practice, candidates can be prioritized by predicted affinity, stability, and energy scores.

Looking ahead, several directions emerge. Addressing data scarcity via PLM-based transfer learning, contrastive learning on mirrored representations, and generative augmentation could improve generalization to D-peptides. Incorporating energy-based refinement or differentiable docking into training may better enforce biophysical realism. Finally, extending D-Flow to multi-objective optimization—balancing stability, affinity, and pharmacological properties—could enable a more general and adaptive peptide design framework.

\section{Conclusion}
\label{Conclusion}
This work introduces D-Flow, which generates D-peptides from scratch. It produces peptides containing a considerable proportion of D-residues through a novel mirror-image algorithm. In addition, it incorporates an adapter-guided protein language model with structural awareness and employs a ControlNet-style mechanism to bridge the gap between pretraining and fine-tuning stages. By utilizing a mirror-image adaptation mechanism, D-Flow successfully generates pure D-peptides, achieving comparable or superior stability and binding affinity relative to their L-peptide counterparts. This represents a significant advancement in generative peptide design and demonstrates strong performance across key metrics.

\section*{References}
\bibliographystyle{IEEEtran}
\bibliography{cite}

\end{document}